\documentclass[
    ,final            
  ]
  {aipproc}

\layoutstyle{6x9}

\begin{document}
\title{Sub-barrier Fusion Cross Sections with Energy Density Formalism}


\author{Muhammad Zamrun F., K. Hagino, N. Takigawa}{
address={Department of Physics, Tohoku University, 980-8578 Japan}
}

\begin{abstract}
We discuss the applicability of 
the energy density formalism (EDF) for 
heavy-ion fusion reactions at sub-barrier energies. 
For this purpose, we calculate 
the fusion excitation function
and the fusion barrier distribution for 
the reactions of 
$^{16}$O with $^{154,}$$^{144}$Sm,$^{186}$W and
$^{208}$Pb with the coupled-channels method. 
We also discuss 
the effect of saturation property on the fusion cross section 
for the reaction between two $^{64}$Ni nuclei, 
in connection to the so called steep fall-off phenomenon of fusion
cross sections at deep sub-barrier energies.
\end{abstract}

\maketitle

\section{Introduction}

The internuclear potential is one of the most important ingredients in 
describing heavy-ion reactions. 
The double folding model (DFM) [1,2] has been widely used 
in order to construct it microscopically. 
This model uses the sudden approximation  
i.e., the assumption that the density of each colliding ion is kept at all
distances during the collision. 
The effect of nucleon exchange between the colliding nuclei 
is partly taken into account in this model through the so called 
knock-on exchange potential. 
Since it does not fully include the exchange effect, the saturation 
property of nuclear matter is respected only partly. 

Since heavy-ion fusion reactions probe the region inside the Coulomb barrier,
where the projectile and target nuclei appreciably overlap with each
other, the effect of saturation plays an important role
[3]. 
There is actually a model for internuclear potential which consistently 
takes account of the saturation property of nuclear matter. 
That is the energy density formalism (EDF) [4-10], 
firstly proposed by Brueckner et al. [4]. 
This model constructs the internuclear potential 
from an energy functional for a dinucleus system. 
Earlier studies have shown that this method 
can account for the elastic scattering of 
$^{16}$O+$^{16}$O reaction [4] and 
the experimental barrier height for many systems [5]. 
Brink and Stancu have investigated intensively the applicability of 
this method using the Skyrme energy functional [6-8]. 
They also showed that the EDF potential is consistent 
with the proximity potential. A similar conclusion was also obtained
in Ref. [9]  using a higher-order Thomas-Fermi approximation for
the kinetic energy and spin orbit densities. 
More recently, the EDF was applied 
to the simplified coupled-channels calculations for
heavy-ion fusion reaction at sub-barrier energies [10]. 

In this contribution, we 
apply the EDF to heavy-ion fusion reactions and perform 
the full order coupled-channels calculations. 
In particular, we analyze the fusion reactions of 
$^{16}$O with
$^{154,}$$^{144}$Sm,$^{186}$W and $^{208}$Pb. 
We also discuss 
the effect of saturation property on the fusion cross section at  
energies close to, and  well below, the Coulomb barrier for
$^{64}$Ni+$^{64}$Ni fusion reaction, for which the 
so called steep fall-off phenomenon was recently reported [11]. 

\section{Energy Density Formalism}

In the energy density formalism, the internuclear potential is assumed
to be given by an energy density for the dinuclear system consisting
of the target and projectile nuclei. 
If one takes 
the frozen density approximation, 
it is given as 
\begin{eqnarray}
V(R)=\int \{\varepsilon[\rho^{(P)}_p(\vec{r})+
  \rho^{(T)}_p(\vec{r},\vec{R}),\rho^{(P)}_n(\vec{r})+
  \rho^{(T)}_n(\vec{r},\vec{R})]\qquad\quad\quad\nonumber \\
 -\varepsilon[\rho^{(P)}_p(\vec{r}),\rho^{(P)}_n(\vec{r})]
  -\varepsilon[\rho^{(T)}_p(\vec{r},\vec{R}),\rho^{(T)}_n(\vec{r},\vec{R})]\}\,\, d\vec{r}.   
\end{eqnarray}
Here $\varepsilon[\rho_{p}(\vec{r}),\rho_{n}(\vec{r})]$
  is the energy density functional, and $\rho^{(P,T)}_p(\vec{r})$ and
  $\rho^{(P,T)}_n(\vec{r})$ are the proton (p) and neutron (n) density
  distributions of the projectile (P) and 
  target (T) nuclei, respectively. 
$\rho(\vec{r},\vec{R})$ represents the density whose center is 
at $\vec{R}$. 
The first term in eq.(1)
  represents the total energy of the system when two 
ions are separated by distance $R$, while the second and the third
terms are the ground state energy of each ion.
In this contribution, we use the Skyrme functional for 
$\varepsilon[\rho_{p}(\vec{r}),\rho_{n}(\vec{r})]$. 
See Refs. [6,12,13] for its explicit form. 

We estimate 
the kinetic energy
and spin orbit densities in the 
semi-classical 
extended Thomas-Fermi approximation [9,14]. 
In this way, the internuclear potential is entirely determined by 
the density
distributions for the colliding nuclei. 
We evaluate them with the Skyrme-Hartree-Fock (SHF) method 
using the same parameter set of the Skyrme interaction 
as that we employ for 
calculating the internuclear potential. 
The pairing correlation is taken into account 
in the BCS approximation with the 
constant gap approach. We take 
$\Delta_p=\Delta_n=11.2/\sqrt{A}$ for this purpose. 
We fit the SHF density 
with a modified Fermi function in evaluating  
the internuclear potential according to Eq.(1). 
We introduce an 
overall scaling factor to the potential obtained in this way so as to
reproduce the experimental data.

\section{Results and Discussions}

\subsection{Coupled-channels calculations with EDF}

We now apply the EDF to 
the reactions of $^{16}$O with
$^{154,}$$^{144}$Sm,$^{186}$W and $^{208}$Pb. 
In the following calculations, we 
use the SkM* parameter set [15]. 
This parameter set gives the incompressibility of nuclear matter 
which is close to the experimental value [14]
and has been successfully used for the description of  
ground state properties for many nuclei. 

The channel coupling does not play so important role 
at energies above the Coulomb barrier. 
We therefore first perform the single-channel calculation for each system  
by ignoring nuclear intrinsic excitations and determine   
an overall normalization factor 
of the EDF potential in order to 
reproduce the experimental fusion cross sections at high 
energies. 
In order to facilitate the coupled-channels calculations, which are 
essential at energies below the Coulomb barrier, 
we simulate the surface region of the resultant potential by
Woods-Saxon form.  
The normalization
factor $(N)$, the optimum Woods-Saxon parameters
$(V_0,\,\,r_0,\,\,a)$, and the  corresponding Coulomb 
barrier height $(V_B)$ for each
system are summarized in Table 1. 
We notice from Table 1 that the EDF potential provides 
the surface diffuseness parameter $a$ of 
around \mbox{0.7 fm}, 
which is similar to the result
of double folding model [2] and is almost 
independent of the system.

In performing the coupled channels calculations, 
we introduce the excitation operator for the intrinsic excitation, 
through the radius parameter 
of the target nucleus in the standard way. 
We used the computer code CCFULL [16] 
for numerical calculations.
\begin{table}[hpt]
\begin{tabular}{lccccc}
\hline
\tablehead{1}{c}{b}{System}
  & \tablehead{1}{c}{b}{$N$}  
 & \tablehead{1}{r}{b}{$V_0$ (MeV)}
  & \tablehead{1}{c}{b}{$r_0$ (fm)}
  & \tablehead{1}{c}{b}{$a$ (fm)}
  & \tablehead{1}{c}{b}{$V_B$ (MeV)} \\
\hline
$^{16}$O+$^{144}$Sm & 1.07 & 66.57 & 1.140 & 0.74  &  61.73\\
$^{16}$O+$^{154}$Sm & 1.31 & 82.66 & 1.144 & 0.75  &  59.54\\
$^{16}$O+$^{186}$W  & 1.37 & 86.86 & 1.152 & 0.73  &  69.02\\
$^{16}$O+$^{208}$Pb & 1.47 & 95.20 & 1.150 & 0.74  &  74.73\\
\hline
\end{tabular}
\caption{Normalization factor and optimum Woods-Saxon parameters for
  the EDF potential
for the $^{16}$O+$^{144,154}$Sm, $^{186}$W, and $^{208}$Pb 
reactions.}
\end{table}
\begin{figure}[hp]
  \includegraphics[height=0.34\textheight]{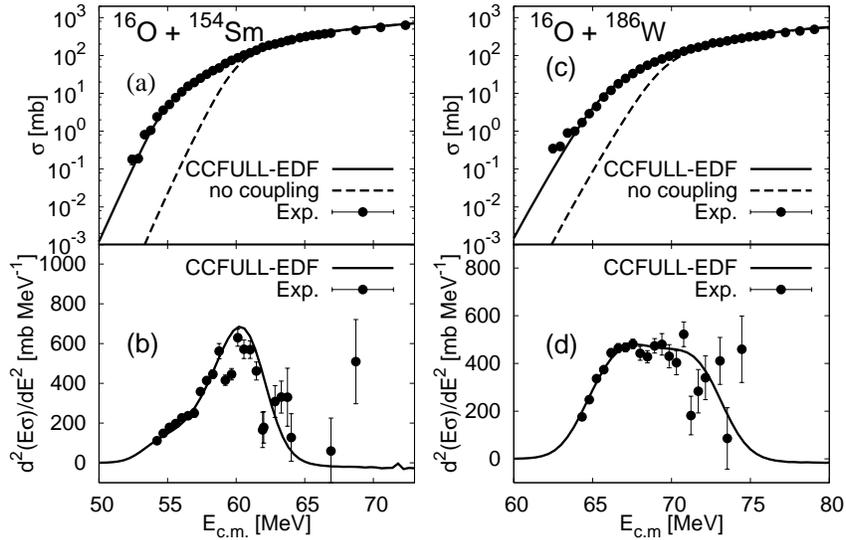}
\caption{Fusion cross sections and fusion barrier distributions 
for the $^{16}$O+$^{154}$Sm (1(a) and 1(b)) and $^{16}$O+$^{186}$W
  (1(c) and 1(d)) reactions. Experimental data are taken from
Ref. [18].} 
\end{figure}

Figures 1(a) and 1(c) show the fusion cross sections for
$^{16}$O+$^{154}$Sm and $^{186}$W reactions, respectively, as 
functions of the incident energy in the center of mass frame. 
The corresponding fusion barrier distributions are shown 
in \mbox{Figs. 1(b) and
1(d).} We include the deformation 
parameters up-to $\beta_6$ of the target nucleus in both cases 
[17]. 
The ground state rotational band up-to the
$10^+$ and $14^+$ member of the $^{154}$Sm and 
$^{186}$W, respectively, is taken into account.
We determine the deformation parameters 
by fitting to the experimental fusion cross sections. 
The resultant 
deformation parameters are $\beta_2=0.33$, $\beta_4=0.035$ and
$\beta_6=0.033$ for $^{154}$Sm, and 
$\beta_2=0.335$, $\beta_4=-0.045$, and $\beta_6=0.018$ 
for $^{186}$W. 
These values are similar to those obtained in [17]. 
The figure clearly shows that our calculations well reproduce
the experimental data. 
\begin{figure}[hbp]
  \includegraphics[height=.34\textheight]{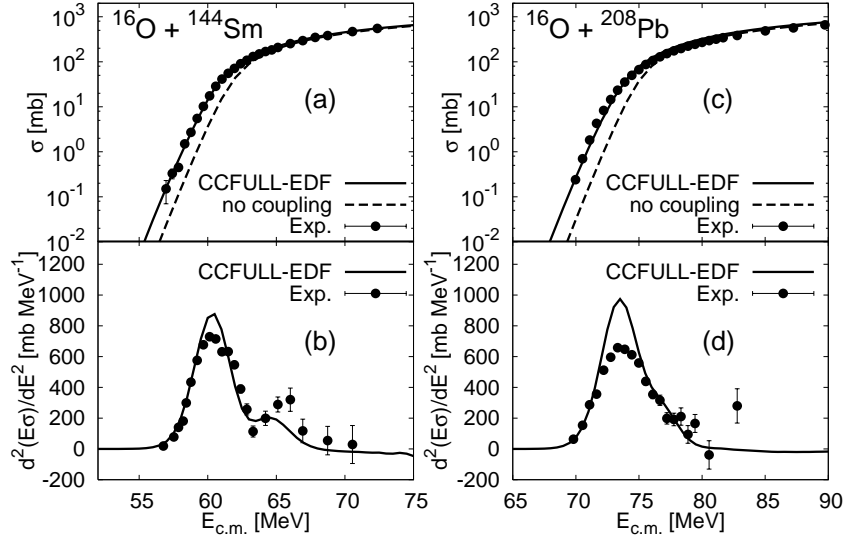}
  \caption{
Same as Fig. 1, but for 
$^{16}$O+$^{144}$Sm (2(a) and 2(b)) and 
$^{16}$O+$^{208}$Pb (2(c) and 2(d)) reactions. The experimental 
  data are taken from Refs. [18,19]. }
\end{figure}\\
We next study the fusion reactions with 
spherical target nuclei, that is  
$^{16}$O+$^{144}$Sm and $^{208}$Pb reactions. 
We include the couplings to the 
$2^+$ and $3^-$ vibrational states in $^{144}$Sm and to the 
$3^-$ and $5^-$ states in
$^{208}$Pb. 
We estimate the deformation parameters from the experimental 
B(E2), B(E3) and B(E5) values.
The excitation energies and  
deformation parameters are $E_2=1.66$ MeV,
$\beta_2=0.11$ and \mbox{$E_3=1.81$ MeV,} $\beta_3=0.205$  for  $^{144}$Sm
and  $E_3=2.615$ MeV, $\beta_3=0.161$ and \mbox{$E_5=3.928$ MeV,}
 $\beta_5=0.056$ for $^{208}$Pb. 
The results of the coupled channels calculations are 
compared with the experimental data in Fig. 2.
We see again that the present calculations 
well reproduce the
experimental data of the fusion cross sections for both 
systems.  

\subsection{Effect of incompressibility}

We next discuss the effect of incompressibility of nuclear matter
on the fusion cross section. We are especially interested in the 
connection between the nuclear incompressibility and 
the steep fall-off problem at deep sub-barrier energies. 
We therefore choose the fusion reactions of two $^{64}$Ni nuclei, 
whose fusion excitation function shows the steep fall-off problem [11]. 
Figures 3(a) and 3(b) show 
the total potential and the nuclear potential for this system 
obtained with EDF using 
three different Skyrme parameters. 
The solid, dashed, and dotted lines have been obtained with 
SIII ($K_{\infty}$=355.4 MeV) [20], 
\mbox{SGI ($K_{\infty}$=269 MeV) [21],} and 
SkM* ($K_{\infty}$=216.7 MeV) [15] parameter sets, 
respectively. 
One observes in Fig. 3(b) that the nuclear potential 
tends to be shallower and more repulsive with increasing incompressibility. 
The fusion excitation function slightly reflects these differences 
as shown in Figs. 3(c) and 3(d). 

A more important observation in connection with 
the steep fall-off problem is that the nuclear potential, 
hence also the total potential, have a much shallower depth at the 
potential minimum compared to the corresponding potentials given by 
the double folding model (DFM) irrespective to the choice of 
the force parameters. The DFM using the 
M3Y force and the same densities as those in the present EDF 
does not actually show a potential pocket, and the depth is 
as large as $- 2500$ MeV and $- 2250$ MeV for the nuclear and total 
potentials, respectively. The EDF using the Skyrme 
force yields a shallow potential irrespective to the parameter sets, 
because the nuclear saturation property is taken into account 
to some extent for all of them. 
Interestingly, as shown in Fig.3(a), 
the minimum energy of the potential pocket nearly equals to that 
discussed by Misicu and Esbensen [3], 
who modified the DFM 
by adding a repulsive term in order to explain the steep-fall off phenomenon. 
The minimum position, which is about 80 MeV in the 
present calculation, is comparable to energy 
$E_S\sim$ 87 MeV, where 
the data of the fusion excitation function start to fall steeply. 
\begin{figure}[htp]
  \includegraphics[height=.33\textheight]{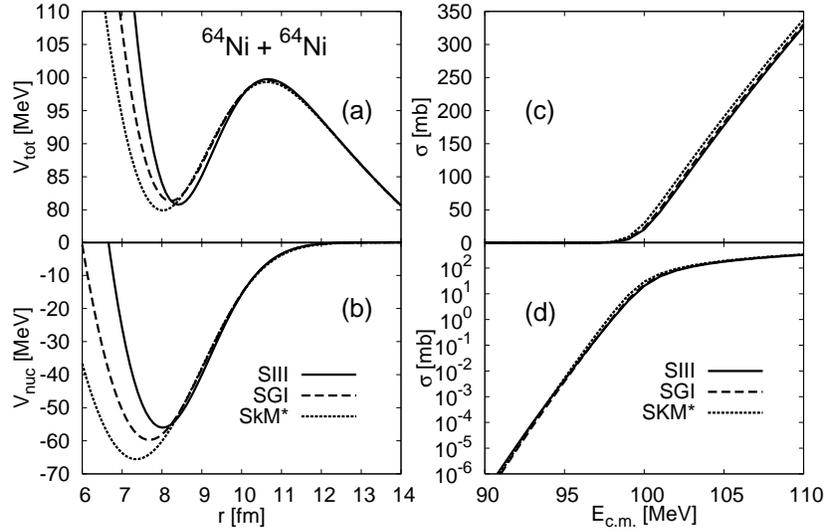}
  \caption{The total (Fig. (3a)) and nuclear (Fig. (3b)) potentials 
for  $^{64}$Ni+ $^{64}$Ni reaction calculated
  with three different Skyrme forces.
Figs. 3(c) and 3(d) show the 
fusion cross sections obtained with these potentials.}
\end{figure}
\section{Conclusions and Further Perspectives}

We have performed the coupled-channels calculations based on the 
EDF for \mbox{$^{16}$O + $^{154,144}$Sm,$^{186}$W and $^{208}$Pb} reactions.  
We have shown that our calculations 
reproduce well the 
experimental data of the fusion excitation function as well as the fusion
barrier distribution for these systems. 
Also, using $^{64}$Ni+$^{64}$Ni reactions, we have shown that the 
EDF potential given by the Skyrme energy density has a much shallower 
depth than the standard DFM and suggested 
that the nuclear saturation property may provide 
an origin of the steep fall-off phenomenon at deep sub-barrier energies. 

In the present studies, we employed the frozen density
approximation, where the total density of the system is simply given 
by the sum of the densities of the projectile and target nuclei. 
This approximation leads to the unphysically high density matter when
two colliding nuclei completely overlap and may break down inside 
the Coulomb barrier. 
This problem can be, at least partly, resolved by respecting 
the Pauli principle, i.e. the role of antisymetrization in the
calculation of the densities of two colliding nuclei [6,22].
Another problem which should be examined is the adiabaticity of the fusion
reactions. The frozen density approximation implies that 
the reaction takes place suddenly. However, it is not obvious whether 
the sudden approach holds to a good approximation for the reactions
at low energies. The opposite limit is the adiabatic approximation, 
where the densities of the  colliding ions change dynamically at every
instant. The EDF can accommodate both 
limits in a natural way, and is suited to examine the
adiabaticity of the 
reactions. 
In connection with the steep fall-off problem, it is an 
interesting question to see at what energy the present 
sudden approximation breaks down and whether the potential minimum 
still remains shallow  
even if one goes beyond the sudden approximation. 
A work towards these directions is now in progress. 

\begin{theacknowledgments}

  This work was partly supported by The 21st Century Center of
  Excellence Program ``Exploring New Science by Bridging
  Particle-Matter Hierarchy'' of the Tohoku University, 
and Monbukagakusho Scholarship from the Japanese Ministry of
  Education, Culture, Sports, Science and Technology.
This work was also supported by the Grant-in-Aid for Scientific Research,
Contract No. 16740139 from the Japanese Ministry of Education,
Culture, Sports, Science, and Technology.

\end{theacknowledgments}


\begin{thebibliography}{99}

\bibitem{1} G.R. Satchler and W.G. Love, \emph{Phys. Rep.} 
\textbf{55}, 183 (1979).   
\bibitem{2}  I.I. Gontchar, D.J. Hinde, M.Dasgupta, and
  J.O. Newton, \emph{Phys. Rev. C}\textbf{69}, 024610 (2004).
\bibitem{3}S. Misicu and H. Esbensen, \emph{Phys. Rev. Lett.} 
\textbf{96}, 112701 (2006).
\bibitem{4} K.A. Brueckner, J.R. Buchler and M.M. Kelly,
  \emph{Phys. Rev.} \textbf{173}, 944 (1968).
\bibitem{5} C. Ngo et al., \emph{Nucl. Phys. A}\textbf{240}, 353 (1975).
\bibitem{6} D.M. Brink and Fl. Stancu,
  \emph{Nucl. Phys. A}\textbf{243}, 175 (1975).
\bibitem{7}Fl. Stancu and  D.M. Brink,
  \emph{Nucl. Phys. A}\textbf{270}, 236 (1976).
\bibitem{8}  D.M. Brink and Fl. Stancu
  \emph{Nucl. Phys. A}\textbf{299}, 321 (1978).
\bibitem{9} A. Dobrowolski, K. Pomorski and J. Bartel,
 \emph{Nucl. Phys. A}\textbf{729}, 713 (2003).
\bibitem{10} Min Liu, et al., \emph{Nucl. Phys A}\textbf{768}, 80 (2006).
\bibitem{11}C.L. Jiang, et al., \emph{Phys. Rev. Lett.} \textbf{93},
  012701 (2004); \emph{Prog. Theor. Phys. Suppl.}
  \textbf{154}, 61 (2004).
\bibitem{12} D. Vautherin and D.M. Brink,
  \emph{Phys. Rev. C}\textbf{5}, 626 (1972).
\bibitem{13} J. Bartel and K. Bencheikh,
  \emph{Eur. Phys. J. A}\textbf{14}, 179 (2002).
 \bibitem{14} M. Brack, C. Guet and H.B. Hakanson,
 \emph{Phys. Rep.} \textbf{123}, 275 (1985).
 \bibitem{15} J. Bartel, P. Quentin, M. Brack, C. Guet and
  H.B. Hakanson, \emph{Nucl. Phys. A}\textbf{386}, 79 (1982).
\bibitem{16} K. Hagino, N. Rowley and A.T. Kruppa,
  \emph{ Comput. Phys. Commun.} \textbf{123}, 143 (1999).
\bibitem{17}Tamanna Rumin, K. Hagino and N. Takigawa,
  \emph{Phys. Rev. C}\textbf{61}, 014605 (1999).
\bibitem{18}J.R. Leigh et al.,  \emph{Phys. Rev. C}\textbf{52}, 3151 (1995).
\bibitem{19}C.R. Morton et al., \emph{Phys. Rev. C}\textbf{60},
  044608 (1999). 
\bibitem{20} M. Beiner, H. Flocard, Nguyen Van Giai and P. Quentin,
  \emph{Nucl. Phys. A}\textbf{238}, 29 (1975).
\bibitem{21} Nguyen Van Giai and H. Sagawa,
  \emph{Phys. Lett. B}\textbf{106}, 379 (1981). 
\bibitem{22}K. Hagino and K. Washiyama, contribution to this conference,
 e-print: nucl-th/0605017.

\end{thebibliography}
\end{document}